\documentclass[aps,prl,twocolumn,superscriptaddress,showpacs]{revtex4} 
\usepackage[dvips]{graphicx} 
\usepackage{latexsym} 
 
\begin{document}

\title{Field-angle-dependent specific heat measurements and  
gap determination of a heavy fermion superconductor 
URu$_2$Si$_2$ 
}

\author{ 
K. Yano 
} 
\affiliation{ 
Institute for Solid State Physics, University of Tokyo, Kashiwa, 
Chiba 270-8581, 
 Japan 
} 
\author{ 
T. Sakakibara 
} 
\affiliation{ 
Institute for Solid State Physics, University of Tokyo, Kashiwa, 
Chiba 270-8581, 
Japan 
} 
\author{ 
T. Tayama 
} 
\affiliation{ 
Institute for Solid State Physics, University of Tokyo, Kashiwa, 
Chiba 270-8581, 
Japan 
} 
\author{ 
M. Yokoyama 
} 
\affiliation{ 
Faculty of Science, Ibaraki University, Mito 310-8512, Japan 
} 
\author{ 
H. Amitsuka 
} 
\affiliation{ 
Graduate School of Science, Hokkaido University, Sapporo 060-0810, Japan 
} 
\author{ 
Y. Homma  
} 
\affiliation{ 
IMR Tohoku University,  Oarai 311-1313, Japan 
} 
\author{ 
P. Miranovi\'{c} 
} 
\affiliation{ 
Department of Physics, University of Montenegro, Podgorica 81000, Montenegro 
} 
\author{ 
M. Ichioka 
} 
\affiliation{ 
Department of Physics, Okayama University, Okayama 700-8530, Japan 
} 
\author{ 
Y. Tsutsumi 
} 
\affiliation{ 
Department of Physics, Okayama University, Okayama 700-8530, Japan 
} 
\author{ 
K. Machida 
} 
\affiliation{ 
Department of Physics, Okayama University, Okayama 700-8530, Japan 
}

\date{\today} 
 
\begin{abstract} 
To identify the superconducting gap structure in URu$_2$Si$_2$ we perform  
field-angle-dependent specific heat measurements for the
two principal orientations in addition to field rotations, 
and theoretical analysis based on microscopic calculations.  
The Sommerfeld coefficient $\gamma(H)$'s in the mixed state 
exhibit distinctively different field-dependence. 
This comes from point nodes and 
substantial Pauli paramagnetic effect of URu$_2$Si$_2$. 
These two features combined give rise to a consistent picture 
of superconducting properties, including a possible first order 
transition of $H_{\rm c2}$ at low temperatures. 
\end{abstract}

 
\maketitle

It is believed that heavy fermion superconductors (SCs), because of their heavy  
effective mass arising from strong electron correlation effects, 
mostly exhibit unconventional pairing states other 
than a $s$-wave pairing to avoid strong on-site repulsion. 
Among various known materials  
URu$_2$Si$_2$ is relatively old, discovered in 1985~\cite{palstra}, 
and has yet full of mysteries. A phase transition at $T_{\rm o}$=17.5~K  
which was thought to be an antiferromagnetic order with 
a tiny moment ~\cite{broholm}, is now  
under lively debate on its origin~\cite{amitsuka}. 
Under this so-called ``hidden order'' (HO) 
the superconducting state appears at $T_{\rm c}=1.3$~K. 
 
The specific heat $C(T)\propto T^2$ at low $T$~\cite{fisher,brison2} 
and nuclear relaxation rate $T_1^{-1}\propto T^3$~\cite{kohori}
suggest a line node gap or more accurately that  
the density of states (DOS) at low energy $N(E)\propto |E|$. 
Up to now, further details of the pairing symmetry in URu$_2$Si$_2$ remain 
unknown: Where the line node is if any, or other alternatives. 

Usually by a bulk thermodynamic measurement alone, 
it is impossible  to locate the position of 
nodes on the Fermi surface (FS), 
except for using the field-angle-dependent 
methods~\cite{vekhter,miranovic2,miranovic,adachi,Sakakibara07,Matsuda06}. 
For example, $d_{xy}$ and $d_{x^2-y^2}$-wave gap structures can 
be distinguished by rotating an external field $H$ relative 
to crystal axis. 
The maximum (minimum) of the oscillation amplitude in the Sommerfeld 
coefficient $\gamma(H)$ in low-$T$ specific 
heat~\cite{miranovic2,miranovic, Sakakibara07} or thermal conductivity $\kappa(H)/T$~\cite{Matsuda06}
corresponds to the anti-nodal (nodal) direction~\cite{kappa}. 
Note that 
${\rm lim}_{T\rightarrow 0}C(H)/T\!=\!\gamma(H)\!=\!{2\over 3}\pi^2\hbar^2N(0,H)$,  
where $N(0,H)$ is the zero energy DOS (ZEDOS) at the Fermi level. 

\begin{figure}[b]
\includegraphics[width=8cm]{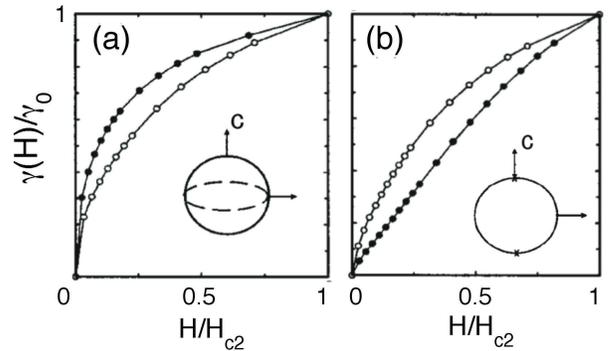}
  \caption{
$H$-dependence of low-$T$ $\gamma(H)$
for $H \parallel c$ ($\bullet$) and $H \parallel a$ ($\circ$), 
calculated by quasiclassical theory~\cite{miranovic2}.
We plot $\gamma(H)/\gamma_0$ vs $H/H_{c2}$ 
for the line node on the equator (a) and point nodes on the poles (b), 
as  schematically shown there. 
$\gamma_0$ denotes the Sommerfeld coefficient in the normal state. 
}
 \label{fig-miranovic}
\end{figure}
In this Letter, using the field-angle-dependent specific 
heat measurement,  
we try to clarify the gap structure 
of superconductivity in URu$_2$Si$_2$. 
For the identification of the point nodes located at north and 
south poles on the FS in tetragonal crystal, 
simple measurement by the field rotation is not enough, 
because the twofold oscillation pattern 
due to large tetragonal anisotropy 
of the upper critical field $H_{\rm c2}$ 
hinders the gap structure. 
Therefore, as an extension of the angle-dependent methods, 
we examine the $H$ dependence of $\gamma(H)$ for each field orientation, 
as shown in Fig.  \ref{fig-miranovic}. 
To identify the point nodes, we use a specialty of them~\cite{miranovic2}. 
As seen in Fig. \ref{fig-miranovic}(b), 
$\gamma(H)$ behaves differently on $H\!\parallel\! c$ and $H\!\parallel\! a$, 
because main contributions to $\gamma(H)$ come from the excited 
quasiparticles (QPs) with the Fermi momentum perpendicular to $H$.  
Namely, $\gamma_c(H)\propto H$ for $H\!\parallel\! c$ 
where the existing point nodes are not sensed 
by the excited QPs, yielding a slow rise in $\gamma_c(H)$. 
This linear $\gamma_c(H)$ 
resembles that expected for the full gap structure~\cite{nakai}.
On the other hand, $\gamma_a(H)$ for $H\!\parallel\! a$ shows a 
steep increase at lower $H$,  
because  the point nodes are effectively sensed 
by the excited QPs for this $H$ direction. 
This $\gamma_a(H)$ behavior is similar to the so-called
Volovik $\sqrt H$ dependence~\cite{volovik}.
This feature is absent in the line node case 
in Fig. \ref{fig-miranovic}(a), 
because the excited QPs feel more 
or less the nodal structure for both $H$ orientations, leading to essentially the same behavior $\gamma_{a,c}(H)\propto \sqrt H$  with slightly different coefficient. 

Through our analysis,  
it will also become clear that the 
Pauli paramagnetic depairing effect is operative for both directions
in URu$_2$Si$_2$.  
Previously two experiments point out this fact; 
a strong depression of $H_{c2}$~\cite{brison} and 
decreasing tendency of the Maki parameter $\kappa_2$ 
upon cooling for both directions~\cite{tenya1,tenya2}. 
The following experiment and analyses reinforce this, 
suggesting a possible first order phase transition at lower $T$~\cite{matsuda}.

\begin{figure}[t]
\includegraphics[width=8.5cm]{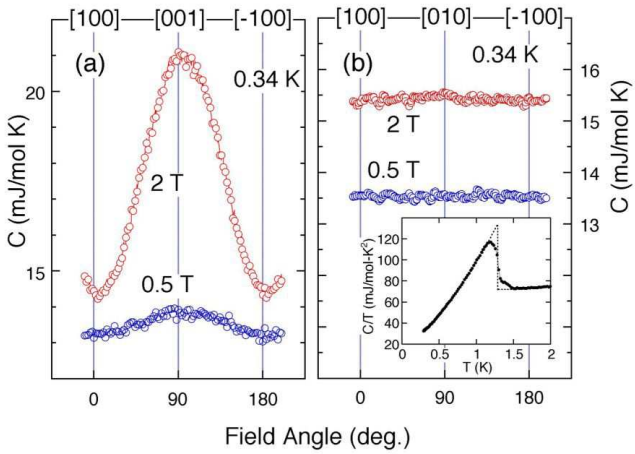}
  \caption{
(Color online) 
$C(H, \theta)$ of URu$_2$Si$_2$ at $T=0.34$~K.
The field-rotational plane is  perpendicular to [010] (a)
and  to [001] (b) directions.
Inset shows $C/T$ vs. $T$ plot at $H$=0.
}
 \label{fig1}
\end{figure}

{\it Experiment.} 
We have carried out the angle-dependent specific heat measurements
$C(H, \theta)$ on a single crystal URu$_2$Si$_2$ (100~mg weight), 
grown by the Czochralski method and annealed 
in high vacuum for 7 days at 950$^\circ$C. 
The crystal had $T_{\rm c}$ of 1.3~K as shown in the inset of Fig. \ref{fig1}. 
$C(H, \theta)$ measurements were carried out by means of a semi-adiabatic method in horizontal fields up to 7~T at a lower $T$ of 0.34~K~\cite{Sakakibara07}. 

In Fig. \ref{fig1} the data $C(H, \theta)$ are shown under the field rotated 
in the (010) (Fig. \ref{fig1}(a))  and (001) plane (Fig. \ref{fig1}(b)). 
The former exhibits a large 
twofold oscillation, which is mainly due to a large anisotropy 
in $H_{\rm c2}$ ($\mu_0H_{\rm c2}^a\!\simeq\! 13$~T, 
$\mu_0H_{\rm c2}^c\!\simeq\! 3$~T for $T\!\rightarrow\! 0$).
It should be noticed that minima of $C(H, \theta)$ 
in Fig. \ref{fig1}(a) do not correspond to nodal directions.
Unfortunately, any subtle oscillation  of $C(H, \theta
)$ 
in the $ac$-plane that could result from nodal structures is washed out 
by the strong anisotropy in $H_{\rm c2}$. 
Thus, we have to examine the $H$-dependence of $C(H,\theta)$ 
at each field orientation. 

We have also searched for a possible line nodal structure running parallel to the $c$ axis,
which can be detected by rotating $H$ within the (001) plane. 
As shown in Fig. \ref{fig1}(b) there is no apparent oscillation 
within the experimental accuracy. 
We could conclude that there is no line node parallel to the $c$ axis 
because the expected oscillation amplitude 
($3\sim 4 \%$)~\cite{miranovic2,miranovic} is beyond 
the experimental resolution ($\sim 0.3\%$).
However this cannot exclude the possibility for the line node along 
the equator on the FS and weaker gap anisotropy parallel to 
the $c$ axis.  

\begin{figure}[t]
\includegraphics[width=8.5cm]{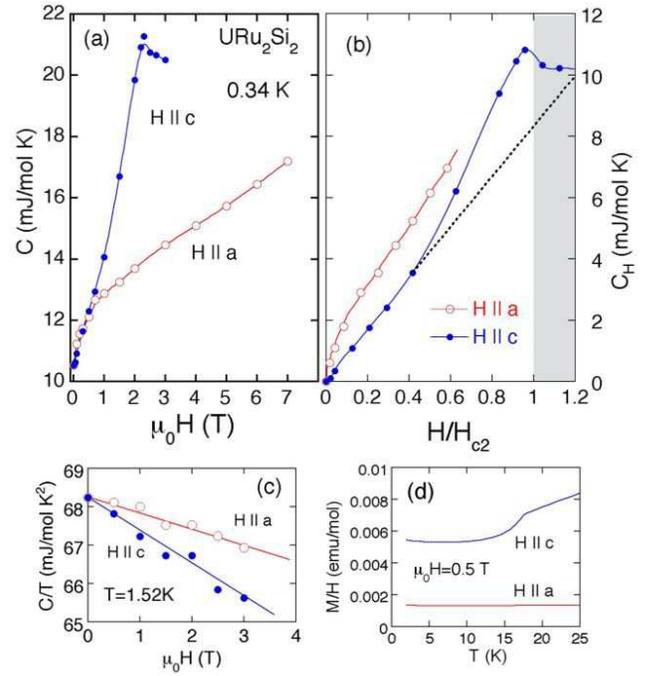}
  \caption{(Color online)
(a) Field variation of $C(H)$ of URu$_2$Si$_2$ at $T\!=\!0.34$~K
for $H\!\parallel\! a$ (open circles)
and $H\!\parallel\! c$ (solid circles). Solid lines are guides 
to the eyes. 
(b) $H$-dependent part of $C$ 
as a function of $H/H_{\rm c2}$. 
The dotted line is a linear extrapolation of the low-field part 
 for $H\!\parallel\! c$.
(c) $H$-variation of $C$ in the normal state 
at 1.52~K for $H\!\parallel\! a$ and $H\!\parallel\! c$.
(d) $T$-variation of $M/H$ in the normal state 
below 25~K for 
$H\!\parallel\! a$ and $H\!\parallel\! c$.
}
 \label{fig2}
\end{figure}

In order to investigate the possible nodal structure, 
we examined the field variation of the specific heat in detail. 
Figure \ref{fig2}(a) displays the results at $T$=0.34~K 
for both $H\!\parallel\! a$ and $H\!\parallel\! c$. 
$C_{\rm c}(H)$ for $H\!\parallel\! c$ peaks at $H_{\rm c2}^c$ ($\sim$2.5~T) 
and reaches the normal-state value at higher $H$. 
For $H\!\parallel\! a$, in contrast, 
$C_{\rm a}(H)$ continues to increase at 7~T 
because $H_{\rm c2}^a$ ($\!\sim\!12.5$~T) is much higher in this direction. 
Accordingly, a significant anisotropy in $C_{\rm a,c}(H)$ develops 
at high $H$.
Surprisingly, this anisotropy in $C_{\rm a,c}(H)$ rapidly diminishes 
with decreasing $H$ and both curves almost fall on top of each other 
below 0.5~T. 
This unexpected behavior can be seen more clearly in Fig. \ref{fig2}(b), 
where the field dependent part of $C$ is plotted 
as a function of $H/H_{\rm c2}$ for the two directions. 
It is remarkable to find the distinctively different field dependence 
of $C_{\rm a,c}(H/H_{\rm c2})$ at low $H$.  
The rapid rise in $C_{\rm a}(H/H_{\rm c2})$ is reminiscent of 
a nodal structure. 
$C_{\rm c}(H/H_{\rm c2})$ shows, in contrast, 
a much weaker linear rise, 
suggesting a full gap feature. 
This distinctive $C(H)$ behavior for two directions 
never occurs when the line nodes exist on the FS, 
as discussed in Fig. \ref{fig-miranovic}(a). 
Thus, the experimental characteristics of URu$_2$Si$_2$ 
in Fig. \ref{fig2}(b) 
are qualitatively consistent with the gap structure 
with the point nodes locating at the north and south poles 
on the FS, as shown in Fig. \ref{fig-miranovic}(b).

Since superconductivity in URu$_2$Si$_2$ coexists 
with HO that sets in at 17.5~K, 
one might suspect that its field-direction dependent 
excitations could result in the anisotropic behavior of  
$C_{\rm a,c}(H/H_{\rm c2})$ at low $H$.  
In order to rule out this possibility, 
we carried out the $C(H)$ measurements  at 1.52~K 
just outside the superconducting state 
but still well inside the HO phase (Fig.~\ref{fig2}(c)). 
For both field directions, 
$C_{\rm a,c}(H)/T$ is only weakly $H$-dependent; 
a slight linear decrease by 2$\sim$3\% has been observed at 2~T.
Since the $C_{\rm a,c}(H)/T$ variation in the superconducting state 
(Fig. \ref{fig2}(a)) is much larger (40$\sim$100\% at 2~T),  
this change in the normal-state background is negligible.
In order to back up this argument, we have also measured the magnetization
$M(T)$ at $\mu_0H$=0.5~T (Fig.~\ref{fig2}(d)).
It is well known that $M_{\rm a}(T)$ for $H\!\parallel\! a$ is nearly 
$T$-independent whereas  
$M_{\rm c}(T)$ for $H\!\parallel\! c$ exhibits a kink upon HO at
17.5~K and rapidly decreases at lower $T$~\cite{palstra}. 
For both directions, $M/H$ below 10~K, 
where HO is well developed, 
is virtually $T$-independent. 
The slight upturn in $M_{\rm a,c}(T)$
seen below 5~K in Fig. \ref{fig2}(d) is presumably due to impurity effect.
Then making use of a Maxwell relation 
$\partial^2 M/\partial T^2=T^{-1}\partial C/\partial H$, 
field variation of $C(H)$ is expected to be small, 
consistent with the results in Fig. \ref{fig2}(c). 
Thus we  conclude that the contribution of HO 
to $C(H)$ is small and negligible; the distinctively different field dependence 
of $C_{\rm a,c}(H/H_{\rm c2})$ in Fig. \ref{fig2}(b) arises 
from superconductivity in the system.

Compared with theoretical curves in Fig. \ref{fig-miranovic}(b),  
$\gamma(H)$ of the experimental data in Fig.~\ref{fig2}(b) is 
suppressed at the middle-$H$ region 
for both directions.  
Especially for $H\!\parallel\! c$, starting from a gradual linear increase 
(dotted line) at low $H$, 
$\gamma(H)$ shows concave curvature at high $H$, 
which is in contrast to the theoretical curve 
in Fig. \ref{fig-miranovic}(b) showing convex curvature.  
We may attribute the anomalous behavior to the Pauli paramagnetic depairing
that suppresses superconductivity eminently at high $H$. Then $H_{\rm c2}$ is reduced,
and $\gamma(H)$ rapidly rises at high $H$.

According to the $M(H)$ measurements~\cite{tenya1,tenya2}, the Maki parameter  
$\kappa_2\!\propto\! 1/\sqrt{(\partial M/\partial H)}_{H=H_{\rm c2}}$ 
is an increasing function of $T$ for both orientations, which is opposite 
to an ordinary SC~\cite{saint}.  
At $H\!=\!H_{\rm c2}$ they show large paramagnetic 
moments build in  the mixed state. 
These facts clearly indicate that the paramagnetic effect is an important 
ingredient in fully understanding this material.

{\it Theoretical calculation}. 
To support above suggestions,   
we calculate the ZEDOS $N(0,H)$ 
by the microscopic quasi-classical theory in the clean limit 
including the paramagnetic effect. 
As is done in Ref.~\cite{ichioka},  
we selfconsistently determine the spatial structure of 
the pair potential $\Delta({\bf r})$ and 
the vector potential ${\bf A}({\bf r})$, 
to appropriately evaluate the vortex core contribution, 
through the quasi-classical Green's functions 
$g( \omega_n, {\bf k},{\bf r})$,  
$f( \omega_n, {\bf k},{\bf r})$, and  
$f^\dagger( \omega_n, {\bf k},{\bf r})$, 
which are calculated by the Eilenberger equation 
\begin{eqnarray} &&  
\left\{ \omega_n +{\rm i} \mu B  
+{\bf k} \cdot\left( \nabla+{\rm i}{\bf A} \right) \right\} f 
=\Delta \phi({\bf k}) g, 
\nonumber  
\\ &&  
\left\{ \omega_n +{\rm i} \mu B
-{\bf k} \cdot\left( \nabla-{\rm i}{\bf A} \right) \right\} f^\dagger 
=\Delta^\ast \phi^\ast({\bf k})g,    \quad  
\label{eq:Eil} 
\end{eqnarray}
with $g=(1-ff^\dagger)^{1/2}$, ${\rm Re}  g > 0$,  
Matsubara frequency $\omega_n=(2n+1)\pi T$, 
and effective Zeeman energy $\mu B$. 
$\mu$ characterizes the strength of the paramagnetic effect.
The self-consistent calculation is performed at $T=0.1T_c$ 
in the triangular vortex lattice 
for ${\bf k}$ on an isotropic Fermi sphere.  
The local DOS at an energy $E$ is given by
$N(E,{\bf r})= \langle 
{\rm Re}\, g({\rm i}\omega_n\to E+{\rm i} 0^+,{\bf k},{\bf r})
\rangle_{\bf k}$ 
with the Fermi sphere average $\langle \cdots \rangle_{\bf k}$. 
By the spatial average of $N(E=0,{\bf r})$, we obtain the ZEDOS   
$N(0,H)= \langle N(E=0,{\bf r}) \rangle_{\bf r}$. 


\begin{figure}[t]
\includegraphics[width=8.0cm]{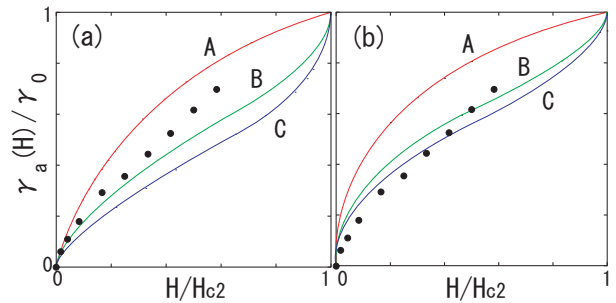}
  \caption{
 (Color online)  
$H$-dependence of $\gamma(H)$ for $H\parallel a$ 
in the linear point node case $\phi({\bf k})=\sin \theta$ (a) 
and the quadratic point node case $\phi({\bf k})=\sin^2 \theta$ (b). 
Solid circles are  the experimental data in Fig. \ref{fig2}(b).  
$\mu=$0 (A), 0.85 (B), 1.7 (C) in (a). 
$\mu=$0 (A), 0.6 (B), 1.0 (C) in (b) 
}
 \label{fig-a}
\end{figure}
\begin{figure}[t] 
\includegraphics[width=8.0cm]{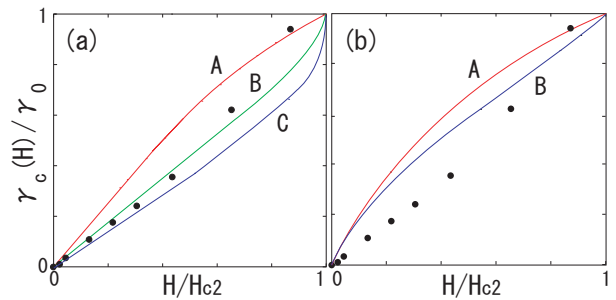}
  \caption{
(Color online) 
The same as in Fig. \ref{fig-a}, but  for $H\parallel c$.  
$\mu=$0 (A), 1.6 (B), 2.5 (C) in (a).
$\mu=$0 (A), 1.3 (B) in (b). 
}
 \label{fig-c}
\end{figure}

In Figs.~4 and 5 we examine two cases of point nodes,   
(i) quadratic point nodes by the pairing function 
$\phi({\bf k})\!=\!\sin ^2\theta$, and 
(ii) linear point nodes by $\phi(\bf k)\!=\!\sin \theta$,   
where $\theta$ is the polar angle from the $c$ axis. 
The former (i) is consistent with the experiments 
$C(T)\propto T^2$~\cite{fisher,brison2} and 
$T_1^{-1}\propto T^3$~\cite{kohori}. 
However, as shown in Fig.~\ref{fig-c}(b), 
the resulting $\gamma_c(H)$ exhibits upward curvature at low $H$,  
which fails to explain the linear $H$-dependence of our experimental data.  
The latter (ii), 
used in the calculation for Fig.~\ref{fig-miranovic}, 
can reproduce $\sqrt{H}$-like behavior  
for $H\!\parallel\! a$ (Fig. \ref{fig-a}(a)) 
and $H$-linear behavior for $H\!\parallel\! c$ (Fig. \ref{fig-c}(a)).  
For better fitting to the experimental data,  
we have to include the paramagnetic effect by $\mu$. 
The best fittings at low $H$ are attained by $\mu_a=0.85$ ($\mu_c=1.60$)
for $H\!\parallel\! a$ ($H\!\parallel\! c$).
The discrepancy at higher $H$ is due to thermal effect
because the experiment was not done at sufficiently low $T$.
Near $T_{\rm c}(H)$, $C(H)/T$ is larger than $\gamma(H)$.
According to the $M(H)$ measurements~\cite{tenya1,tenya2}, 
the paramagnetic moments at $H_{\rm c2}$ are comparable for two directions, 
{\it i.e.} $\mu_a^2H^a_{\rm c2}\simeq \mu_c^2H^c_{\rm c2}$. 
This indicates $\mu_c/\mu_a=2.1$, which is consistent to 
our choice $\mu_c/\mu_a=1.60/0.85\sim1.9$.

Our calculation can show that 
above a critical value $\mu^*\!\sim\! 0.4$
the first order transition occurs at 
$H_{c2}$ in lower $T$ and higher $H$ 
for a Fermi sphere and $s$ wave pairing.  
Thus, our assigned $\mu$ values are within the first order region. 
Since the $C(H)$ data were taken at $T\!=\!0.34$~K,  
there is no indication for it. 
However, recent thermal conductivity measurements~\cite{matsuda} 
clearly shows a jump of $\kappa(H)$ at $H_{\rm c2}$ 
for both directions around $T\sim 0.1$~K.
This is in accord with our assignment for the $\mu$ values. 

Among the possible pairing function in the group theoretical 
classifications for a tetragonal crystal~\cite{gorkov, ozaki}, 
spin-triplet symmetries such as $\tau_xk_x+\tau_yk_y$, 
$\tau_xk_x+i\tau_yk_y$, $\tau_z(k_x+ik_y)$, 
and $(\tau_x+i\tau_y)(k_x+ik_y$)  
have point nodes ($\tau_j=i\sigma_2\sigma_j$),  
but they are unlikely to be realized in URu$_2$Si$_2$
because the paramagnetic effect is present in both directions.
In spin-singlet symmetries, 
$k_x^2+k_y^2$ (A$_{1g}$) has two quadratic point nodes,  
and $k_z(k_x+ik_y)$ (E$_{g}$) consists of two linear point nodes 
and a line node. 
Our analysis supports linear point nodes, 
and excludes line nodes. 
Thus, if we choose $k_z(k_x+ik_y)$ as a 
plausible gap function with linear point nodes, 
the line node contribution may be smeared 
out due to the actual FS topology~\cite{kasahara}. 
While we have to consider realistic FS topology 
for conclusive evaluation of the gap structure, 
the linear point nodes in the polar direction 
are plausible, because low-$H$ behavior of $\gamma(H)$  
is governed by the low-$E$ QPs  excited
around nodes~\cite{volovik}. 

CeCoIn$_5$ is also considered to be a Pauli limited SC with a 
tetragonal structure. The phase diagrams in $H$ vs $T$ for both field directions  
are similar and exhibit Pauli-limited
behavior. It is understandable because $M$ values at  
$H_{\rm c2}$ are almost same for both
$H\!\parallel\! c$ and $H\!\parallel\! a$~\cite{Tayama02}, which  
matters the paramagnetic effect.
This situation is the same as in URu$_2$Si$_2$ as mentioned before. These  
two compounds belong to a clean limit SC
free from dirt effects, such as spin-orbit scattering that tends to  
mask a first order transition
due to the paramagnetic effect.

In summary, 
we performed field-angle-dependent specific heat measurements
in URu$_2$Si$_2$. 
The Sommerfeld coefficient $\gamma(H)$ exhibits distinctively 
different behavior for $H \parallel c$ and $H \parallel a$. 
These reveal, supported by microscopic calculation, 
that the gap structure posses point nodes
at north and south poles on the Fermi surface 
and moreover the Pauli paramagnetic effect is important in this system. 
This also suggests a first order transition 
at $H_{\rm c2}$ in lower temperatures, which has been 
confirmed recently~\cite{matsuda}. 

We thank Y. Matsuda for informative discussions.


%
\bibliography{basename of .bib file}

\begin{thebibliography}{99} 

\bibitem{palstra}

T.T.M. Palstra {\it et al}., Phys. Rev. Lett. {\bf 55}, 2727 (1985);
M.B. Maple {\it et al}., Phys. Rev. Lett. {\bf 56}, 185 (1986);
W. Schlabitz {\it et al}.,  Z. Phys. B {\bf 62}, 171 (1986). 

\bibitem{broholm}
C. Broholm {\it et al}.,  Phys. Rev. Lett. {\bf 58}, 1467 (1987).

\bibitem{amitsuka}
C.R. Wiebe {\it et al}., Nature Phys. {\bf 3}, 96 (2007);
H. Amitsuka {\it et al}., J. Mag. Mag. Mater. {\bf 310}, 214 (2007);
Y.S. Oh {\it et al}., Phys. Rev. Lett. {\bf 98}, 016401 (2007).

\bibitem{fisher}
R.A. Fisher {\it et al}., Physica B {\bf 163}, 419 (1990).

\bibitem{brison2}
J.P. Brison {\it et al}., Physica B {\bf 199-200}, 70 (1994).

\bibitem{kohori}
K. Matsuda {\it et al}., J. Phys. Soc. Jpn. {\bf 65}, 679 (1996).

\bibitem{vekhter}
I. Vekhter {\it et al}., Phys. Rev.  B {\bf 59}, R9023 (1999). 

\bibitem{miranovic2} 
P. Miranovi\'{c}  {\it et al}., Phys. Rev. B {\bf 68}, 052501 (2003). 

\bibitem{miranovic}
P. Miranovi\'{c} {\it et al}., 
J. Phys.: Condens. Matter {\bf 17}, 7971 (2005).

\bibitem{adachi} 
H. Adachi {\it et al}., Phys. Rev. Lett. {\bf 94}, 067007 (2005). 

\bibitem{Sakakibara07}
T. Sakakibara {\it et al}., J. Phys. Soc. Jpn. {\bf 76}, 051004 (2007), and references therein.

\bibitem{Matsuda06}
Y. Matsuda  {\it et al}., J. Phys. Condens. Matter {\bf 18} R705 (2006), and references therein.

\bibitem{kappa}
The interpretation of the oscillation pattern in $\kappa(H)/T$ 
is complicated by the additional factor due to 
the directional dependent scattering mechanism in some case. 

\bibitem{nakai} 
N. Nakai {\it et al}., Phys. Rev. B {\bf 70}, 100503(R) (2004). 

\bibitem{volovik} 
G.E. Volovik, Pis'ma Zh. Eksp. Teor. Fiz. {\bf 58}, 457 (1993) 
[JETP Lett. {\bf 58}, 469 (1993)]. 

\bibitem{brison}
 J.P. Brison, {\it et al},  Physica  C {\bf 250}, 128(1995). 

\bibitem{tenya1}
 K. Tenya {\it et al}., Physica B {\bf 281-282}, 991 (2000).

\bibitem{tenya2}
 K. Tenya {\it et al}.,  preprint. 

\bibitem{matsuda} 
Y. Kasahara {\it et al.}, Phys. Rev. Lett. {\bf 99}, 116402 (2007). 

\bibitem{saint}
 D. Saint-James, G. Sarma, and E.J. Thomas, 
 {\it Type II Superconductivity} (Pergamon, Oxford, 1969),  
Chap. 6. 

\bibitem{ichioka} 
M. Ichioka and K. Machida, Phys. Rev. B {\bf 76}, 064502 (2007). 

\bibitem{gorkov}
G.E. Volovik and L.P. Gor'kov, Zh. Eksp. Teor. Fiz. {\bf 88}, 1412 (1985)
[Sov. Phys. JETP {\bf 61}, 843 (1985)].
 
\bibitem{ozaki}
 M. Ozaki {\it et al}., Prog. Theor. Phys. {\bf 75}, 442 (1986).

\bibitem{kasahara} 
During the preparation of the manuscript, we have learned of a 
paper~\cite{matsuda}, 
which concludes the same gap structure
$k_z(k_x+ik_y)$ by measuring thermal conductivity.
They assign the point (line) nodes on the main (minor) 
FS with heavy (light) mass. These are completely 
consistent with and reinforce our conclusions. 

\bibitem{Tayama02}
 T. Tayama {\it et al.}, Phys. Rev. B {\textbf 65}, 180504(R) (2002).

\end{thebibliography}

\end{document}